\documentstyle[12pt]{article}
\begin{document}
\thispagestyle{empty}
{\it Invited talk at Third Conference
on Nucleon--Antinucleon Physics, Moscow, Russia, September 11--16, 1995}
\vspace{1.5cm}
\begin{center}
\begin{large}
{\bf Variational calculation of antiprotonic helium atoms}\\
\vspace{0.3truecm}
O.~I.~Kartavtsev\\
\end{large}
Bogoliubov Laboratory of Theoretical Physics \\
Joint Institute for Nuclear Research\\
141980, Dubna, Russia\\
\end{center}

\sloppy

\begin{abstract}
A few per cent fraction  of antiprotons stopped in helium survives
 for an enormous time (up to tens of microseconds)
in comparison with the usual lifetime ($10^{-12}s$)
of these particles in matter. The explanation is that
antiprotons  are captured in the metastable antiprotonic helium
atoms $^{3,4}\! H\! e\bar pe$ carrying an extremely large total
angular momentum $L\sim 30-40$. Initial populations,
level lifetimes and very precise values of the transition energies
were obtained in experiments on the resonant laser--induced
annihilation. Analogous long-lived systems were observed in
experiments with negative kaons and pions.

The purpose of this report is to present the results of
calculation of eigenenergies, radiative transition rates,
energy-level splitting due to relativistic interactions and
Auger decay rates of antiprotonic helium atoms within
the variational approach.
\end{abstract}
\newpage
\section{Introduction}

The lifetime of negatively charged heavy particles (such as $\pi^-,
K^-, \bar p$) stopped in matter would be about $10^{-12}s$ due to
nuclear absorption after exotic--atom formation. Recently it has
been found that a small fraction (a few per cent) of kaons~\cite{yam},
 pions~\cite{nakpi} and antiprotons~\cite{iwa}, \cite{nak} stopped
 in helium media survive for an enormous time (up to tens of
microseconds for antiprotons). The reason for this unusual behavior
is the formation of a metastable three--body system consisting a
helium nucleus, an electron and a negative heavy particle. The
prediction of these systems were made as early as thirty years
ago by Condo~\cite{con} to explain a considerable difference of
pion lifetimes in liquid helium and hydrogen. Later on, this
qualitative prediction was supported by the variational
calculations of Russell~\cite{rus}. One can mention also the first
calculation of the antiprotonic helium atom using the
Born-Oppenheimer approximation~\cite{ahl}. Since antiprotons are
stable and lifetime of an antiprotonic helium is the largest among
these exotic systems, the experimental data concern mainly the
antiprotonic helium atom. The modern experiments on the
resonant laser-induced annihilation~\cite{mor,hay} have initiated
thorough investigations of these unusual systems.  Initial
populations, level lifetimes, and very precise values of the
transition energies (with a relative accuracy $< 10^{-5}$) have
been obtained using this remarkable method.

The antiprotonic helium can be considered as a counterpart of the
usual helium atom with one electron replaced by an antiproton. A
large angular momentum $L\sim~(\mu /m_3)^{1/2}$ ($\mu$ is the
reduced mass of the antiproton--nucleus subsystem and $m_3$ is the
electron mass) provides that the antiproton-nucleus and
electron-nucleus distances are approximately equal. At the same
time, this system can be considered as an exotic diatomic molecule,
where one nucleus is negatively charged~\cite{shim}.

While there is a number of processes destroying the antiprotonic
helium atom, the following considerations explain its very long
lifetime. The
most part of an extremely large total angular momentum  $L \sim
35-40$ belongs to the pair of heavy particles and annihilation of
an antiproton is inhibited by a large centrifugal barrier.  The
Auger decay is inhibited due to sufficiently large values
$(\lambda_0\geq 4)$ of the angular momentum of an outgoing electron.
Usual mechanism of the de-excitation by the Stark mixing is not
appropriate for the three-body system due to the lack of
degeneracy.  The collisional de-excitation by surrounding $He$
atoms is suppressed due to the screening of an antiproton by the
electron in an antiprotonic helium.  The only remaining
de-excitation mechanism is multistep dipole radiative transitions,
whose rates of each step are of order $\mu s^{-1}$.  The discussion
of theoretical calculations on antiprotonic helium atoms and
related topics can be found in~\cite{fb}.

After the formation of the antiprotonic helium its time evolution
is determined by the processes of radiative transition and Auger
decay. The precise description of energy
spectra requires to take into account the minor effect of
relativistic interactions.
The spin-dependent part of relativistic interactions gives rise
to splitting of energy levels and sufficiently large splitting
values can be measured experimentally.
The direct variational method provides a possibility in the
framework of one approach to calculate eigenenergies, energy-level
splitting due to relativistic interactions, radiative transition
rates and Auger decay rates of antiprotonic helium atoms. The
results of these calculations are presented and discussed in this
report. The isotopic effect is considered by comparison of the
calculated properties of $^{4}\!  H\!  e\bar pe$ and $^{3}\!  H\!
e\bar pe$.

\section{Variational method}
\label{secvar}

The nonrelativistic Hamiltonian of the antiprotonic helium atom
in the triangular coordinates is

\begin{equation}
H=-\frac{1}{2\mu}\Delta _{\bf r}-\frac{1}{2\mu_{1}}\Delta
_{\mbox{\boldmath$\rho$} }
-\frac{1}{m_1}{{\bf\nabla_r}\cdot{\bf\nabla}_{\mbox{\boldmath$\rho$}}}
-\frac{2}{r}-\frac{2}{\rho }+\frac{1}{|{\bf r}-\mbox{\boldmath$\rho$} |},
\label{ham}
\end{equation}
where $1/\mu=1/m_1+1/m_2,~1/\mu_1=1/m_1+1/m_3$ and
$m_1, {\bf r}_1, m_2, {\bf r}_2, m_3, {\bf r}_3 $ are masses
and coordinates of the helium nucleus, antiproton and electron,
respectively.
Apart from the triangular coordinates
${\bf r} = {\bf r}_2 - {\bf r}_1,
\mbox{\boldmath$\rho$}={\bf r}_3 - {\bf r}_1$,
the Jacobi coordinates
${\bf r} = {\bf r}_2 - {\bf r}_1,
\mbox{\boldmath$\rho$}_1={\bf r}_3 -(m_1 {\bf r}_1- m_2 {\bf
r}_2)/(m_1+ m_2)$,
corresponding momenta
${\bf p}=-i\nabla_{\bf r}$, \
${\bf q}=-i\nabla_{\mbox{\boldmath$\rho$}}$,\
${\bf q}_1=-i\nabla_{{\mbox{\boldmath$\rho$}}_1}$
and angular momenta
${\bf l}=[{\bf rp}]$, \
$\mbox{\boldmath{$\lambda$}}=[\mbox{\boldmath$\rho$} {\bf q}]$,\
${\mbox{\boldmath{$\lambda$}}}_1=
[{\mbox{\boldmath$\rho$}}_1 {\bf q}_1]$ will be used to simplify
the notation.

Since antiprotonic helium atoms are unstable against the decay to the
$H\! e\bar p~+~e$ channel, the variational method cannot be
directly applied to the calculation of energies and wave functions.
Therefore, according to~\cite{kar} the approximate
Hamiltonian $H_{LN}=P_{LN}HP_{LN}$
which is the projection of the hamiltonian $H$ onto
the closed-channel subspace was constructed to calculate the
N-th energy level $E_{LN}$ of the system for the total angular
momentum $L$.  Explicitly $P_{LN}$ is constructed as a projector
onto the subspace of eigenfunctions of relative angular momenta
$\bf l, \mbox{\boldmath$\lambda$} $ (or $\bf l,
{\mbox{\boldmath$\lambda$}}_1$), belonging
to a limited set of $l, \lambda $ (or $l, \lambda_1$) eigenvalues.
While using the Jacobi coordinates this set is
defined by the condition $l>l_0$, where $l_0$ is the largest pair
angular momentum satisfying the inequality $\varepsilon_{l_0}<E_{LN}$
and $\varepsilon_{l_0}$ is the energy of the hydrogen-like ion
$^{3,4}\! He\bar p$ with the angular momentum $l_0$. If the
triangular coordinates are used, this set can be defined by the
condition $\lambda < \lambda_{0}$, where $\lambda_{0}$ is the
smallest angular momentum satisfying the inequality
$\varepsilon_{L-\lambda_0}<E_{LN}$. By definition, $\lambda_{0}$ is
also a multipolarity of the Auger transition, i.~e. the smallest
angular momentum of the outgoing electron.
These conditions describe the natural way to
provide the approximate Hamiltonian $H_{LN}$ to have at least $N$
 eigenvalues below the boundary of the continuous spectrum and the
variational method can be applied to solve the eigenvalue problem
for $N$ lowest states of $H_{LN}$.

It is worthwhile to mention that the $l=L$ projection is the main
part of the wave function and the contribution of $l\ne L$
projection rapidly decreases with increasing $|l-L|$. For this
reason the approximate Hamiltonian $H_{LN}$ provides an accurate
calculation of energies $E_{LN}$ and wave functions $\Psi_{LN}$ for
an metastable states ($\lambda_0 \geq 4$) of antiprotonic helium
atoms.  In fact, the described conditions on $l, \lambda$ are more
restrictive than necessary for an application of the variational
method.  Practically the accuracy of calculation can be improved
taking into account $\lambda > \lambda_{0}$ (or $l < l_0$ in
the case of the Jacobi coordinates) components of the wave
function. While using these components a convergence of the
calculated values with increasing the number of trial functions
will be investigated to provide the reliable results.

\section{Energy levels}

The variational method described in the previous section was
applied to determine eigenfunctions and eigenenergies of the
equation
\begin{equation}
(H_{LN}-E_{LN})\Psi_{LN}=0 .
\label{schr}
\end{equation}
A set of simple
variational trial functions of the form
\begin{equation}
\chi_{nkl\lambda i}^{LM}({\bf r}, \mbox{\boldmath$\rho$})=
{\cal Y}_{l\lambda}^{LM}({\bf\hat r},{\bf\hat{\mbox{\boldmath$\rho$}}})
r^{l+i}\rho^{\lambda} exp(-a_nr-b_k\rho ) ,
\label{trial}
\end{equation}
where ${\cal Y}_{l\lambda}^{LM}({\bf\hat r},{\bf\hat{\mbox{\boldmath$\rho$}}})$
are bispherical harmonics of angular variables,
was used in the calculations. It is essential that
${\cal Y}_{l\lambda}^{LM}({\bf\hat r},{\bf\hat{\mbox{\boldmath$\rho$}}})$
are eigenfunctions of $\bf l, \mbox{\boldmath$\lambda$} $ and
this form of trial functions allows the easy application of the
projection method described in Section~\ref{secvar}.

Up to 1000 trial functions~(\ref{trial}) containing up to 25
bispherical harmonics were used in the calculations. A set of
nonlinear parameters $a_n, b_k$ for simplicity was chosen in the
form $a_n=a_0\alpha^n, b_k=b_0\beta^k$ and the variation of
parameters $a_0,\alpha,b_0,\beta$ was used to minimize energy
values. As it is usual for the variational method, the precision
falls down with increasing the excitation number $N$. To reach more
precise energy values, a set of trial functions was limited by the
condition $l > L-\lambda_{0}$ or even less restrictive condition $l
> L-\lambda_{0}-1$ instead of $\lambda < \lambda_{0}$.
Convergence of the calculated energies with increasing the number
of trial functions was obtained and results are presented in
Table~\ref{taben}.
\begin{table}[ht]
\caption{Calculated energies $E_{LN}(a.u.)$ for five lowest
energy levels of the $^{3,4}\! H\! e\bar pe$ systems in the range
$31\leq L\leq 37$}
\label{taben}
\begin{tabular}{cccccc}
\hline \hline
\multicolumn{6}{c}{$^{4}\! H\! e\bar pe$}\\
\hline \hline
 L & $E_{L1}$ & $E_{L2}$ & $E_{L3}$ & $E_{L4}$ & $E_{L5}$ \\
\hline 32 & -3.3534091 & -3.2272500 & -3.1161769 & -3.0183048 &  \\
\hline 33 & -3.2158963 & -3.1049859 & -3.0075402 & -2.9219136 & -2.8355338 \\
\hline 34 & -3.0931201 & -2.9959130 & -2.9107010 & -2.8359650 & -2.7685835 \\
\hline 35 & -2.9836255 & -2.8988545 & -2.8247010 & -2.7597140 & -2.7007897 \\
\hline 36 & -2.8862911 & -2.8126761 & -2.7483976 & -2.6920660 & -2.6389254 \\
\hline 37 & -2.7999636 & -2.7364076 & -2.6809542 & -2.6320110 & -2.5853018 \\
\hline \hline
\multicolumn{6}{c}{$^{3}\! H\! e\bar pe$}\\
\hline \hline
 L & $E_{L1}$ & $E_{L2}$ & $E_{L3}$ & $E_{L4}$ & $E_{L5}$ \\
\hline 31 & -3.3484555 & -3.2190748 & -3.1056452 & -3.0061026 & \\
\hline 32 & -3.2073388 & -3.0940418 & -2.9949034 & -2.9082665 & -2.8309327 \\
\hline 33 & -3.0817523 & -2.9829571 & -2.8967378 & -2.8214095 & -2.7539447 \\
\hline 34 & -2.9702506 & -2.8844813 & -2.8097982 & -2.7445651 & -2.7007897 \\
\hline 35 & -2.8714926 & -2.7974268 & -2.7330406 & -2.6763097 & -2.6389254 \\
\hline 36 & -2.7843183 & -2.7207232 & -2.6654633 & -2.6161392 & -2.5853018 \\
\hline \hline
\end{tabular}
\end{table}
Some information on this calculation can be found also
in~\cite{kar}. Energies of the $L,N$ states of ${^4}H\! e{\bar p}e$
and $L-1,N$ states of ${^3}H\! e{\bar p}e$  are very close to each
other.

By using the method of laser-induced resonant annihilation, till
now, wavelengths of five transitions $(35,4\rightarrow
34,4)$~\cite{mor}, $(34,3\rightarrow 33,3)$~\cite{maas},
$(35,2\rightarrow 34,4)$, $(34,3\rightarrow 33,5)$,
$(35,3\rightarrow 34,3)$~\cite{wid} in ${^4}H\! e{\bar p}e$
and two transitions $(34,4\rightarrow 33,4)$, $(33,3\rightarrow
32,3)$~\cite{wid} in ${^3}H\! e{\bar p}e$ were measured with high
accuracy. The calculated and experimental
wavelengths are in agreement with an accuracy not worse
 than $5\cdot 10^{-4}$ and more elaborate trial functions are
 needed to reach a higher accuracy in the energy calculation.
Recently, V.~I.~Korobov has obtained more precise spectra and
transition wavelengths of an antiprotonic helium, using the
correlated trial functions in variational calculations~\cite{kor}.

\section{Radiative transitions}

Due to large lifetimes
of metastable states against the Auger decay and collisional
de-excitation the radiative transitions become the most important
to describe the evalution of the antiprotonic helium atom. As far
as only dipole transitions are significant, the total angular
momentum changes by unity in each transition. Thus, the system
looses the angular momentum and energy step-by-step and finally
reaches the state with a large Auger decay rate.

The rate of the dipole transition $LN\rightarrow L_1N_1$
is given by
\begin{equation} w=\frac{4}{3(2L+1)}|\alpha(E_{LN}-E_{L_1N_1}|
)^3|M_{d}|^2 \displaystyle\frac{me^4}{\hbar^3}s^{-1},
\label{eq:tran}
\end{equation}
where $\alpha =e^2/\hbar c$ is the fine structure constant
and the reduced matrix element of the dipole transition is defined
in the form
 \begin{equation}
M_{d}=\langle\Psi_{L_1N_1}||{\bf
r}+{\mbox{\boldmath$\rho$}}||\Psi_{LN}\rangle.
\label{eq:me}
\end{equation}
Variational wave functions $\Psi_{LN}$ and energies $E_{LN}$ have
been obtained as described in the previous section and used
in~(\ref{eq:tran}),~(\ref{eq:me}) to calculate the radiative
transition rates. These
results are presented in Table~\ref{tabrad}.
A convergence of the calculated values
with increasing the number of trial functions provides an estimate
of the relative accuracy on the level of one per cent.
The radiative transition rates calculated in
papers~\cite{shim},~\cite{mo},~\cite{yamoht}
are fairly close to each other and to present results.
The important feature is the predominance of
transitions between states of the same $N$, i.~e. the approximate
conservation of the excitation number in the radiative transitions.
Thus, radiative cascades in an antiprotonic helium
proceed almost independently along the chains of states of fixed
$N$. However, for higher excitation numbers the probabilities of
interchain transitions become more significant and such
transitions will be taken into account for the cascade description.

As in the energy calculation, there is a correspondence of the $L,
N$ state of $^{4}\! H\!  e\bar pe$ and the $L-1, N$ state of
$^{3}\!  H\!  e\bar pe$ and the radiative transition rates of the
corresponding states are almost the same. At the same time,
the $^{3}\!  H\!  e\bar pe$ transition rates systematically exceed
those of $^{4}\!  H\!  e\bar pe$ in accordance with the difference
of experimental lifetimes of the  antiprotonic helium
atoms~\cite{nak}. However, the explanation of this difference
is not simple because the mean lifetime depends also on another
processes and a significant difference
in populations of the corresponding states was found in the recent
experiments for different isotopes~\cite{wid}.
\begin{table}[htb]
\caption{Radiative transition rates $w(10^5s^{-1})$ from the
$LN$ to $L_1N_1$ state of the $^{3,4}\! H\! e\bar pe$ systems.
Only transition rates $w\geq10^4s^{-1}$ are presented.}
\label{tabrad}
\begin{tabular}{l@{$\rightarrow$}rcl@{\ }l@{$\rightarrow$}rcl@{\quad
}l@{$\rightarrow$}rcl@{\
}l@{$\rightarrow$}rc}
\hline\hline
\multicolumn{7}{c}{$^{4}\! H\! e\bar pe$} & &
\multicolumn{7}{c}{$^{3}\! H\! e\bar pe$} \\
\hline\hline
\multicolumn{2}{c}{transition} & $w$ & &
\multicolumn{2}{c}{transition} & $w$ & &
\multicolumn{2}{c}{transition} & $w$ & &
\multicolumn{2}{c}{transition} & $w$ \\
\hline
37,1 & 36,1 & 6.733 & & 35,5 & 34,5 & 3.143 & & 36,1 & 35,1 & 7.255 & & 34,5 &
33,4 & 1.036 \\ \hline
37,2 & 36,1 & 0.159 & & 34,1 & 33,1 & 7.744 & & 36,2 & 35,1 & 0.187 & & 34,5 &
33,5 & 4.532 \\ \hline
37,2 & 36,2 & 5.859 & & 34,2 & 33,2 & 7.554 & & 36,2 & 35,2 & 6.255 & & 33,1 &
32,1 & 8.483 \\ \hline
37,3 & 36,2 & 0.441 & & 34,3 & 33,1 & 0.206 & & 36,3 & 35,2 & 0.511 & & 33,2 &
32,2 & 8.240 \\ \hline
37,3 & 36,3 & 4.907 & & 34,3 & 33,3 & 6.926 & & 36,3 & 35,3 & 5.182 & & 33,3 &
32,1 & 0.301 \\ \hline
37,4 & 36,3 & 0.742 & & 34,4 & 33,1 & 0.288 & & 36,4 & 35,3 & 0.887 & & 33,3 &
32,3 & 7.539 \\ \hline
37,4 & 36,4 & 3.990 & & 34,4 & 33,2 & 0.137 & & 36,4 & 35,4 & 4.135 & & 33,4 &
32,1 & 0.476 \\ \hline
37,5 & 36,4 & 1.218 & & 34,4 & 33,3 & 0.355 & & 36,5 & 35,4 & 1.642 & & 33,4 &
32,2 & 0.205 \\ \hline
37,5 & 36,5 & 2.947 & & 34,4 & 33,4 & 6.073 & & 36,5 & 35,5 & 2.934 & & 33,4 &
32,3 & 0.387 \\ \hline
36,1 & 35,1 & 7.249 & & 34,5 & 33,1 & 0.355 & & 35,1 & 34,1 & 7.846 & & 33,4 &
32,4 & 6.577 \\ \hline
36,2 & 35,2 & 6.531 & & 34,5 & 33,2 & 0.365 & & 35,2 & 34,1 & 0.102 & & 33,5 &
32,1 & 0.748 \\ \hline
36,3 & 35,2 & 0.333 & & 34,5 & 33,4 & 0.728 & & 35,2 & 34,2 & 7.019 & & 33,5 &
32,2 & 0.571 \\ \hline
36,3 & 35,3 & 5.644 & & 34,5 & 33,5 & 4.840 & & 35,3 & 34,2 & 0.384 & & 33,5 &
32,4 & 0.807 \\ \hline
36,4 & 35,3 & 0.637 & & 33,1 & 32,1 & 7.720 & & 35,3 & 34,3 & 5.995 & &
\multicolumn{2}{c}{} & \multicolumn{1}{c}{} \\ \hline
36,4 & 35,4 & 4.685 & & 33,2 & 32,2 & 7.859 & & 35,4 & 34,3 & 0.747 & &
\multicolumn{2}{c}{} & \multicolumn{1}{c}{} \\ \hline
36,5 & 35,4 & 1.077 & & 33,3 & 32,1 & 0.648 & & 35,4 & 34,4 & 4.937 & &
\multicolumn{2}{c}{} & \multicolumn{1}{c}{} \\ \hline
36,5 & 35,5 & 3.323 & & 33,3 & 32,3 & 7.492 & & 35,5 & 34,4 & 1.304 & &
\multicolumn{2}{c}{} & \multicolumn{1}{c}{} \\ \hline
35,1 & 34,1 & 7.629 & & 33,4 & 32,1 & 1.166 & & 35,5 & 34,5 & 3.666 & &
\multicolumn{2}{c}{} & \multicolumn{1}{c}{} \\ \hline
35,2 & 34,2 & 7.123 & & 33,4 & 32,2 & 0.570 & & 34,1 & 33,1 & 8.276 & &
\multicolumn{2}{c}{} & \multicolumn{1}{c}{} \\ \hline
35,3 & 34,2 & 0.205 & & 33,4 & 32,3 & 0.151 & & 34,2 & 33,2 & 7.700 & &
\multicolumn{2}{c}{} & \multicolumn{1}{c}{} \\ \hline
35,3 & 34,3 & 6.331 & & 33,4 & 32,4 & 6.798 & & 34,3 & 33,2 & 0.236 & &
\multicolumn{2}{c}{} & \multicolumn{1}{c}{} \\ \hline
35,4 & 34,1 & 0.137 & & 33,5 & 32,1 & 3.503 & & 34,3 & 33,3 & 6.797 & &
\multicolumn{2}{c}{} & \multicolumn{1}{c}{} \\ \hline
35,4 & 34,3 & 0.509 & & 33,5 & 32,2 & 1.537 & & 34,4 & 33,1 & 0.169 & &
\multicolumn{2}{c}{} & \multicolumn{1}{c}{} \\ \hline
35,4 & 34,4 & 5.327 & & 33,5 & 32,3 & 0.321 & & 34,4 & 33,3 & 0.582 & &
\multicolumn{2}{c}{} & \multicolumn{1}{c}{} \\ \hline
35,5 & 34,1 & 0.168 & & 33,5 & 32,4 & 0.434 & & 34,4 & 33,4 & 5.762 & &
\multicolumn{2}{c}{} & \multicolumn{1}{c}{} \\ \hline
35,5 & 34,2 & 0.187 & & 33,5 & 32,5 & 0.730 & & 34,5 & 33,1 & 0.199 & &
\multicolumn{2}{c}{} & \multicolumn{1}{c}{} \\ \hline
35,5 & 34,4 & 1.059 & & \multicolumn{2}{c}{} & \multicolumn{1}{c}{} & & 34,5 &
33,2 & 0.244 & & \multicolumn{2}{c}{} & \multicolumn{1}{c}{} \\
\hline\hline
\end{tabular}
\end{table}
\clearpage

Lifetimes of the $(L,N)=(35,4)$ and
$(L,N)=(34,3)$ states of $^{4}\!  He\bar pe$ system were determined
experimentally by the method of resonant laser--induced
annihilation~\cite{hay},~\cite{maas}. Table~\ref{tabrad2} contains the
experimental decay rates and theoretical radiative transition rates
of these states.
\begin{table}[htb]
\caption{Experimental decay rate and theoretical radiative
 transition rates $(10^6 s^{-1})$ for two states of the $^{4}\!
 H\! e\bar pe$ system}
\label{tabrad2}
\begin{tabular}{ccccc} \hline\hline $L,N$ & Experiment &
\cite{mo} & \cite{shim} & present \\ \hline
 $ 35,4 $ & $0.72\pm 0.02$ \cite{hay} & 0.614 & 0.619 & 0.597 \\
 \hline
 $ 34,3 $ & $1.18\pm 0.04$ \cite{maas} & 0.734 & 0.754 &  0.713 \\
 \hline  \hline
 \end{tabular}
 \end{table}
One of the most probable reasons for the difference of
experimental and all the theoretical values indicates the
substantial contribution of additional nonradiative decay channels.
As a density dependence of the lifetime of the (34,3) state
in $^{4}\! H\! e\bar pe$ has been found in recent
experiments~\cite{wid}, collisions with surrounding atoms
contribute to the decay processes. An extrapolation of the new
experimental data to zero density gives $0.91\pm 0.15\mu s^{-1}$,
which is close to the theoretical radiative transition rate.  The
density effect probably takes place also for the 35,4 state and
the remaining discrepancy may be caused by an additional decay
process, e.~g.  the Auger decay.

\section{Energy--level splitting due to relativistic interactions}

The precise measurement
of transition energies of antiprotonic helium atoms in recent
experiments on the laser-induced resonant annihilation~\cite{mor},
{}~\cite{hay} invokes the theoretical description of
energy spectra with a comparable accuracy. That description of
energy spectra requires to take into account the relativistic
corrections of an order of $\alpha ^2$ to the pure Coulomb
interaction.

Since the contribution to energies from relativistic interactions
depends on the antiproton mass, charge and magnetic moment,
these calculations can be used for the precise determination of the
antiproton properties.  This knowledge is essential in testing
the fundamental symmetry principles~\cite{eades}.

The spin-dependent part of the relativistic interactions gives rise
to splitting of energy levels, and each single transition turns into
a multiplet. Sufficiently large distances between lines in the
multiplet can be measured experimentally. It is worthwhile to
mention that the resolution in current experiments is about 10GHz
and without much difficulty can be improved to 1GHz~\cite{pps}.
As it will be discussed below, due to the interaction with electron
spin, antiprotonic helium energy levels split into two multiplets
and the interaction with nuclei spins provides a minor splitting
within each multiplet. Calculation of the former large splitting is
discussed in this section. More details of this calculation were
presented in~\cite{exan}.

For each pair of particles $i, j$ in the three-body system the
relativistic correction of an order of $\alpha ^2$ to the pure
Coulomb two-body potential can be described by the Breit interaction
$U_{ij}^{(B)}$.
The correction to
the kinetic energy of an order of $\alpha ^2$ for each particle $i$
is
\begin{equation}
\Delta T_{i}=-\frac{\alpha
^2}{8}\frac{p_i^4}{m_i^3} \label{kin}
\end{equation}
Full relativistic correction $H_r$ of an order of $\alpha ^2$ to the
three-body nonrelativistic Hamiltonian is a sum of $U_{ij}^{(B)}$ for all
pairs of particles and $\Delta T_i$ for all particles
\begin{equation}
H_{r}=\sum_{i}\Delta T_i+\sum_{i>j}U_{ij}^{(B)}.
\label{relint}
\end{equation}
In the definition of the expressions $U_{ij}^{(B)},
\Delta T_i$ in eq.~(\ref{relint}) it is proposed
that particles momenta ${\bf p}_i$
will be taken in the center of mass system of three particles~\cite{lev}.

The interaction  $H_r$ given in (\ref{relint}) conserves the sum
${\bf J}={\bf L}+\sum_{i}{\bf s}_i$ of
 the total angular momentum $\bf L={\bf l}+\mbox{\boldmath$\lambda$}$
and particle spins ${\bf s}_i$. Each level of the
nonrelativistic Hamiltonian splits into four and eight sublevels
for $^{4}\! H\! e\bar pe$ and $^{3}\! H\! e\bar pe$ systems, respectively.
Due to
very small mass ratios $m_3/m_1, m_3/m_2$, the largest contribution to
the energy splitting comes from the interaction with the electron
spin ${\bf s}_3$.
Taking into consideration only terms responsible for the splitting
in (\ref{relint}),
this part of relativistic interaction can be written as follows:
\begin{eqnarray}
\label{split}
H_s=\alpha^2\bigl(\frac{1}{\rho^{3}}\mbox{\boldmath$\lambda$}{\bf s}_3+
\frac{1}{2|{\bf r}-\mbox{\boldmath$\rho$}|^3}
[{\bf r}-\mbox{\boldmath$\rho$},{\bf q}]{\bf s}_3-\\
\frac{1}{m_2|{\bf r}-\mbox{\boldmath$\rho$}|^3}
[{\bf r}-\mbox{\boldmath$\rho$},{\bf p}]{\bf s}_3+
\frac{2}{m_1\rho^3}[\mbox{\boldmath$\rho$},{\bf p}]{\bf s}_3\bigr)
\nonumber
\end{eqnarray}
While the last two terms in
(\ref{split}) are inversely proportional to the
masses of heavy particle  $m_{1,2}$, their contribution to the energy splitting
is
nevertheless comparable to the contribution from the first two
terms for the following reasons.  The small mass factor is
compensated in part due to the large angular momentum $l \sim L$ of
heavy particles.  At the same time, only small components of the
wave function corresponding to the nonzero electron angular momenta
$\lambda \ne 0$ lead to a nonzero splitting value from the first
two terms in (\ref{split}).


The interaction  $H_s$, given in (\ref{split}), conserves the sum
${\bf j}={\bf L}+{\bf s}_3$ of
 the total angular momentum $\bf L$
and electron spin ${\bf s}_3$ and
splits each level into two sublevels, corresponding to the
eigenvalues $j=L\pm 1/2$.
The part of the interaction depending on heavy particle
spins removes the remaining degeneracy and splits each $j=L\pm 1/2$
sublevel further into two or four levels for the $^{4}\! H\! e\bar
pe$ and $^{3}\! H\! e\bar pe$ systems, respectively. Values of
this secondary splitting are much smaller in comparison with the
splitting arisen due to the interaction with the electron
spin~(\ref{split}).

Since the splitting is small in comparison with energy
differences between states of different $L$ values, the energy
shift $\Delta_{jLN}$ for the state with quantum numbers $j,L,N$
can be found in the first order of perturbation theory in $H_{s}$
\begin{equation}
\Delta_{jLN}=\langle \Psi_{jLN}|H_s|\Psi_{jLN}\rangle,
\label{shift}
\end{equation}
where the wave function $\Psi_{jLN}$ of the $j,L,N$ state is the
vector production of the nonrelativistic wave function $\Psi_{LN}$
and spin function describing the dependence of the electron spin.
Level splitting $\Delta E_{LN}=\Delta_{L+1/2LN}-\Delta_{L-1/2LN}$
is a difference of shifts~(\ref{shift}) for $j=L\pm 1/2$.

Due to smallness of the relativistic interaction, radiative
transitions proceed only between states of the same $j$. For this
reason, each spectral line of the transition
from the state $L_iN_i$ to state $L_fN_f$ is to be split into a
doublet with the interline distance $\Delta\nu =\Delta
E_{L_iN_i}-\Delta E_{L_fN_f}$.


Splitting values $\Delta E_{LN}$
for a number of states of the $^{3,4}\! He\bar pe$
systems in the range of experimentally observed values of the total
angular momentum $L$ have been calculated as described above
by using variational nonrelativistic wave functions $\Psi_{LN}$.
These values are presented in Table~\ref{tab1}.
\begin{table}[ht]
\caption{Splitting values $\Delta E_{LN}$ ($10^{-6}$au) of
the lowest levels in the $^{3,4}\! H\! e\bar pe$ systems.}
\label{tab1}
\begin{tabular}{lllllll} \hline\hline
\multicolumn{7}{c}{$^{4}\! H\! e\bar pe$}\\ \hline
N & L=32 & L=33 & L=34  & L=35  & L=36  &  L=37 \\ \hline
1 & -1.10 & -1.15 & -1.15 & -1.14 & -1.12 & -1.09 \\ \hline
2 & -1.12 & -1.09 & -1.08 & -1.07 & -1.04 & -1.00 \\ \hline
3 & -1.01 & -1.02 & -1.00 & -0.98 & -0.94 & -0.90 \\ \hline
4 &       & -0.94 & -0.94 & -0.90 & -0.86 & -0.82 \\ \hline
5 &       &       & -0.93 & -0.90 & -0.84 & -0.81 \\ \hline \hline
\multicolumn{7}{c}{$^{3}\! H\! e\bar pe$}\\ \hline
N & L=31  & L=32  & L=33  & L=34  & L=35  & L=36 \\ \hline
1 & -1.20 & -1.16 & -1.19 & -1.19 & -1.18 & -1.14 \\ \hline
2 & -1.14 & -1.19 & -1.15 & -1.12 & -1.08 & -1.04 \\ \hline
3 & -1.08 & -1.06 & -1.05 & -1.04 & -1.00 & -0.98 \\ \hline
4 &       & -0.97 & -0.92 & -0.86 & -0.85 & -0.81 \\ \hline
\hline
\end{tabular}
\end{table}

As it follows from expression~(\ref{split}), the form of the wave
function at small interparticle distances is the most important
 in evaluating the integral~(\ref{shift}). Convergence of the
calculated splitting values $\Delta E_{LN}$  provides a few per cent
relative accuracy. It is worthwhile to mention
that due to a better
description of the small $N$ states in the variational method,
the accuracy of calculation for
the $L,N$ state increases with decreasing $N$.
At the same time an accuracy of calculation
decreases with decreasing $L$ due to decreasing the multipolarity
$\lambda_0$ of the Auger decay for large $L$.  Moreover, it is
impossible to trace the convergence in the case of short-lived
states due to a small multipolarity $\lambda_0<3$ of the Auger
decay. This problem is closely connected with a large natural width
of these states which exceeds significantly a splitting value.
Also, the variational procedure meets some difficulties in
describing the short range behavior of the wave function for states
with large enough $N$, especially, in the $^{3}\!  H\! e\bar
pe$ system. These are the reasons to omit the above--mentioned
cases in Table~\ref{tab1}.

The last two terms in eq.~(\ref{split}) describe the interaction of the
electron magnetic moment with the magnetic field of heavy particles.
These terms give rise to the largest contribution to the energy--level
splitting. For a better understanding of the splitting dependence
on $L,N$, this contribution is presented in Table~\ref{tab2} for
the $^{4}\!  H\! e\bar pe$ system. The contribution to the
energy--level splitting from the first two terms in~(\ref{split})
is of opposite sign and smaller in magnitude. Nevertheless,
this contribution decreases with increasing $L$ and compensates
the $L$ dependence of the last two terms in eq.~(\ref{split})
providing a very slow dependence of the total splitting $\Delta
E_{LN}$ on $L$.
\begin{table}[ht]
\caption{Contribution of the last two terms in
$N_s$ to the energy--level splitting $\Delta E_{LN}$
($10^{-6}$au) in the $^{4}\! H\! e\bar pe$ system.}
\label{tab2}
\begin{tabular}{lllllll} \hline\hline
N & L=32  & L=33  & L=34  & L=35  & L=36  &  L=37 \\ \hline
1 & -1.41 & -1.43 & -1.40 & -1.37 & -1.34 & -1.28 \\ \hline
2 & -1.39 & -1.34 & -1.30 & -1.27 & -1.22 & -1.16 \\ \hline
3 & -1.26 & -1.24 & -1.20 & -1.15 & -1.10 & -1.04 \\ \hline
4 &       & -1.14 & -1.11 & -1.06 & -1.00 & -0.94 \\ \hline
5 &       &       & -1.10 & -1.06 & -0.98 & -0.95 \\ \hline   \hline
\end{tabular}
\end{table}
Due to almost exact conservation of the $j$ value in the radiative
transition the spectral line splitting will be found as a difference
of $\Delta E_{LN}$ presented in Table~\ref{tab1}.
Most appropriate for the experimental measurement are
the favoured transitions between states of the same
$N$, which have the largest
radiative rates~\cite{kar},~\cite{yamoht},~\cite{shim}. However,
the calculated splitting values are almost independent of $L$ for a
given $N$, and it is not plausible to resolve such a small
difference in splitting for the favoured transitions.  For this
reason, the experimental proposal for the nearest future~\cite{pps}
is aimed at searching for the splitting in unfavoured  transitions
$(L,N) \to (L-1,N+2)$.

To measure splitting in experiments on the laser--induced
resonant annihilation, the initial state should be long--lived.
This is provided by the condition that the multipolarity of the
Auger decay for this state is $\lambda_0=4$. The next condition is
that the natural width of the short--lived final state will be
smaller than the splitting value, and the multipolarity of the
Auger decay for this state will be $\lambda_0=3$.  The spectral
line splitting $\Delta\nu$ for a number of suitable transitions is
presented in Table~\ref{tab3}.
\begin{table}[ht]
\caption{Spectral line splitting
$\Delta\nu =\Delta E_{L_iN_i}-\Delta E_{L_fN_f}$ (GHz) for the
transitions $L_iN_i \rightarrow L_fN_f$
in the $^{3,4}\! H\! e\bar pe$ systems.}
\label{tab3}
\begin{tabular}{r@{$\rightarrow$}lc|r@{$\rightarrow$}lc} \hline\hline
\multicolumn{3}{c|}{$^{4}\! H\! e\bar pe$} &
\multicolumn{3}{c}{$^{3}\! H\! e\bar pe$}\\ \hline
$L_iN_i$ & $L_fN_f$ & $\Delta\nu$ & $L_iN_i$ & $L_fN_f$ & $\Delta\nu$
\\ \hline
33,1 & 32,3 & -0.92 & 32,1  & 31,3  & -0.53 \\ \hline
34,1 & 33,3 & -0.86 & 33,1  & 32,3  & -0.86 \\ \hline
34,2 & 33,4 & -0.91 & 33,2  & 32,4  & -1.22 \\ \hline
35,2 & 34,4 & -0.87 & 34,2  & 33,4  & -1.35 \\ \hline
35,3 & 34,5 & -0.34 & \multicolumn{3}{c}{}  \\ \hline
\hline
\end{tabular}
\end{table}
These values are of an order of the
experimentally measurable value $\sim $1GHz.  One of the recently
discovered~\cite{wid} unfavoured transitions $(35,2\rightarrow
34,4)$ in ${^4}H\! e{\bar p}e$ is a good candidate for the
splitting measurement.

Energy--level splitting $\Delta E_{LN}$ decreases with increasing
the excitation number $N$ and an obvious reason for this
dependence is decreasing of the wave function at small
interparticle distances for excited states. One can mention an
appreciable difference in the $\Delta E_{LN}$ dependence on $N$ for
the $^{4}\! H\! e\bar pe$ and $^{3}\! H\! e\bar pe$ systems.  As a
consequence, an appreciable isotopic effect appears also for
the spectral line splitting $\Delta\nu$.

The following considerations can be used to understand
qualitatively the $L,N$--dependence of the energy--level
splitting. The contribution to splitting from
the interaction of the electron magnetic
moment with the magnetic field of heavy particles is described by
the last two terms in the splitting interaction $H_s$~(\ref{split}).
This contribution is
proportional to the relative momentum of heavy particles $p$.
One can consider that the motion of heavy particles is
approximately the same as in a hydrogen--like atom and the momentum
$p$ is inversely proportional to the angular momentum $L$. This is
the reason for increasing this contribution with decreasing $L$, as
presented in Table~\ref{tab2}.  The contribution from the first two
terms in the splitting interaction $H_s$ is connected with the
electron rotation and is proportional to a small component of the
wave function arising due to polarization of an electron by an
antiproton. With decreasing $L$ the antiproton moves to the
region of increasing electron density and the polarization
increases. In this way, contributions to the energy--level
splitting from the last two terms in $H_s$ and the remaining part
of splitting interaction are of opposite sign and level off the
dependence of the total splitting $\Delta E_{LN}$ on $L$.

One can consider quasi-classically that the antiproton orbit
became more stretched with increasing $N$ at a fixed
total angular momentum. For this reason all the terms of the
splitting interaction $H_s$ decrease with increasing $L$ and provide
the $N$ dependence presented in Tables~\ref{tab1},~\ref{tab3}.

\section{Auger decay}

In addition to the radiative transitions, the
important decay channel of antiprotonic helium atoms is  the
emission of an electron with the $^{3,4}\! He\bar p$ hydrogen-like
ion formation, i.~e. the Auger decay.  The main feature of the
Auger decay rates of antiprotonic helium atoms is the essential
dependence on the transition multipolarity, i.~e.  the smallest
angular momentum of the outgoing electron $\lambda_0$. As discussed
in Section~\ref{secvar}, calculated eigenenergies
$E_{LN}$~\cite{shim},~\cite{yamoht},~\cite{kar} unambiguously
determine the transition multipolarities $\lambda_0$ by the
condition $\varepsilon_{L-\lambda_0}<E_{LN}$.

Apart from the early calculation of Russell~\cite{rus} the only
progress in the calculation of the Auger decay rates is due to the
paper~\cite{mo}.  It was found that the Auger decay rate decreases
about three orders of magnitude with increasing the smallest
angular momentum of the outgoing electron $\lambda_0$ by one.

As the Auger lifetime of states with $\lambda_0=3$ and
$\lambda_0=4$ is of an order of $10^{-8} s$ and $10^{-5} s$,
respectively, and the radiative lifetime is about $10^{-6}
s$~\cite{shim},~\cite{yamoht},~\cite{kar}, the resonant
laser-induced transitions between states of $\lambda_0=3$ and $4$
became observable in the experiments ~\cite{mor},~\cite{hay}. For
this reason, the most important thing is to calculate the Auger
decay rates for the states of $^{3,4}\! He\bar p e$ systems with
multipolarities $\lambda_0=3,4$. The method and results of the
calculation are presented in~\cite{aug} and will be described in
this section. One should mention that there are no other
calculations of the Auger decay rate for the $^{3}\!  He\bar pe$
system and the presented results allow one to consider the isotopic
effect in this process.

The Auger decay rates of the antiprotonic helium atom  are very
small in the atomic scale and the very fine details of the
discrete state wave functions $\Psi_{LN}$ and continuum wave
functions $\Psi_c$ will be accurately determined. In
particular, since $\Psi_{LN}$ and $\Psi_c$ belong to
the closed- and open-channel subspaces, respectively, the
orthogonality condition $\langle\Psi_{LN} | \Psi_c\rangle=0$ will
be strictly fulfilled. As described in Section~\ref{secvar},
$\Psi_{LN}$ is an eigenfunction of the approximate Hamiltonian
$H_{LN}$ and projector $P_{LN}$. As a consequence, $\Psi_c$ should
be an eigenfunction of the projector $1-P_{LN}$ and can be found
as a solution of the equation
\begin{equation}
(1-P_{LN})H(1-P_{LN})\Psi_c=E_{LN}\Psi_c.
\label{eqpsic}
\end{equation}
Thus, the orthogonality requirement is fulfilled and the
perturbation theory can be applied due to smallness of the decay
rates.

According to the Feshbach orthogonal projection method
the decay rate $\Lambda$ is
\begin{equation}
\Lambda=\frac{1}{\sqrt{2 \mu_3 (E -\varepsilon_{l_0}) }} |M_{tr}|^2
\frac{m_3 e^4}{\hbar^3} \mbox{s}^{-1},
\label{Lambda}
\end{equation}
where $1/\mu_3=1/(m_1 +m_2)+ 1/m_3 $ and the transition matrix
element is
\begin{equation}
M_{tr}=\langle \Psi_c |(1-P_{LN}) H P_{LN}|\Psi_{LN} \rangle.
\label{tme}
\end{equation}
Since the continuum wave function is naturally described in the
Jacobi coordinates ${\bf r},\mbox{\boldmath$\rho$}_1$, these
coordinates will be used in this calculation.

The contribution to $M_{tr}$ is negligible for the
$\Psi_c$ components, corresponding to $(l, \lambda_1)$
eigenvalues of angular momenta $\bf l, \mbox{\boldmath$\lambda$}_1$
 if $l\ne l_0, \lambda_1 \ne \lambda_0$. For this reason, only the
$(l_0, \lambda_0)$ component will be taken into account in the
calculation of $\Psi_c$.

Due to the large centrifugal barrier the continuum wave function $\Psi_c$
can be taken as a product of the normalized
wave function of $^{3,4}\! He\bar p$ and the function $f(\rho_1$)
describing the relative motion of an electron and $^{3,4}\! He\bar
p$.  As the total angular momentum of the system is $L$ and its
projection is $M$, $\Psi_c$ takes the form
\begin{equation}
\label{con}
\Psi_c({\bf r},\mbox{\boldmath$\rho$}_1)=A\; {\cal Y}^{LM}_{l_0
\lambda_0}(\hat{\bf r},\hat{\mbox{\boldmath$\rho$}}_1)\; r^{l_0} \;
e^{-a r}\;f(\rho_1)=\Phi({\bf
r},\hat{\mbox{\boldmath$\rho$}}_1)f(\rho_1).
\end{equation}
In the
framework of this approach the interaction of the outgoing electron
and the remaining $^{3,4}\! He\bar p$ ion is described by the
folding potential
\begin{equation} V_0(\rho_1)=\langle
\Phi|(1-P_{LN})H(1-P_{LN})|\Phi\rangle,
\end{equation}
and the radial function $f(\rho_1)$ obeys the ordinary
differential equation.

Due to the large centrifugal barrier for the outgoing electron, the
Auger decay rate is
mainly determined by the wave functions in the large $\rho_1$ range, where
the interaction of
the electron and $^{3,4}\! He\bar p$ system
is nearly a sum of the Coulomb and centrifugal potential.
Indeed, the replacement of the folding potential by the Coulomb one
gives rise to a minor change in the matrix element. This fact
supports the applicability of the approximation~(\ref{con}).


500--650 trial functions $\chi_{nkl\lambda i}^{LM}({\bf r},
\mbox{\boldmath$\rho$}_1)$~(\ref{trial}) of the variables ${\bf r},
\mbox{\boldmath$\rho$}_1$ have been used in the variational
calculation of $\Psi_{LN}, E_{LN}$.
As described above, to fulfil the strict requirement of
orthogonality, only trial functions satisfying the condition
$l>l_0$ have been taken into account.  The largest contributions to
the transition matrix element $M_{tr}$ come from the ($l,L-l$)
components of the wave function and they have been treated with
special care. An additional set of 70--130 trial functions with
specific nonlinear parameters have been used to describe these
components.  As the calculated energy values are in good agreement
with the results of~\cite{kar}, one can conclude that the present
calculations provide the accurate description of antiprotonic
helium atoms.  Using $\Psi_{LN}, E_{LN}$ and $\Psi_c({\bf r})$ in
the form~(\ref{con}) the Auger decay rates $\Lambda$ have been
calculated according to equations~(\ref{Lambda}),~(\ref{tme}).
These results are presented and compared with~\cite{mo} in
Table~\ref{tabaug}.
\begin{table}[ht]
\caption{Multipolarities $\lambda_0$ and Auger decay rates $\Lambda
$ ($s^{-1}$) of $^{3,4}\rm He\bar{p} e$} \label{tabaug}
\begin{minipage}{\textwidth}
\renewcommand{\thefootnote}{\fnsymbol{footnote}}
\renewcommand{\thempfootnote}{\fnsymbol{mpfootnote}}
\begin{tabular}{cccc|ccc} \hline\hline \multicolumn{4}{c}{$^{4}\rm
He\bar pe$} & \multicolumn{3}{c}{$^{3}\rm He\bar pe$} \\
\hline\hline
 L,N & $\lambda_0$ & $\Lambda$ & $\Lambda$~\cite{mo} & L,N & $\lambda_0$ &
$\Lambda$\\
\hline
 34,4& 3 &$7 \cdot 10^7$   &$8.5 \cdot 10^7$     & 33,4 & 3 & $2 \cdot 10^8$
\\
\hline
 33,3& 3 &$1.2 \cdot 10^8$   & $1.5 \cdot 10^8$\footnote[1]{K.~Ohtsuki, private
communication} & 32,3 & 3 & $3.4 \cdot 10^8$\\
\hline
 32,2& 3 & $1.1 \cdot 10^8 $ & $4.0 \cdot 10^7$\footnotemark[1] & 31,2 & 3 &
$3.3 \cdot 10^8$\\
 \hline 35,4& 4 &$\sim 5 \cdot 10^4$ & $6.7 \cdot  10^2$ & 34,4 & 4 & $\sim
10^5$\\
 \hline 34,3& 4 &$\sim 1  \cdot 10^5$ &$2.7 \cdot 10^4$  & 33,3 & 4 & $ \sim
10^5$\\
 \hline\hline
\end{tabular}
\end{minipage}
\end{table}
To understand the role of the wave function structure in
this calculation, the largest contributions of the ($l,\lambda_1$)
components of $\Psi_{LN}$
to the transition matrix element are presented in Table~\ref{tabaug3}.
It is worthwhile to mention that the exponentially small part of
the wave function in the large $\rho_1$ region is important
in evaluating the integral~(\ref{tme}).
\begin{table}[ht]
\caption{Normalized contributions to the transition matrix element
$M_{tr}(l,\lambda_1)$ from the ($l,\lambda_1$)
components of the wave function for two states of the
$\rm {}^4 He \bar{p}e$ system.}
\label{tabaug3}
\begin{tabular}{ cc|cc}
\hline\hline
\multicolumn{2}{l|}{(L,N)=(34,3) $\lambda_0$=4} &
\multicolumn{2}{l}{(L,N)=(34,4) $\lambda_0$=3}\\
\hline
$(l,\lambda_1)$ & $M_{tr}(l,\lambda_1)$ &
$(l,\lambda_1)$ & $M_{tr}(l,\lambda_1)$ \\
\hline
34,0 & -0.28 & 34,0 & 0.24 \\ \hline
33,1 &  0.51 & 33,1 & -0.46 \\ \hline
32,2 & -0.38 & 32,2 &  1.22 \\ \hline
31,3 &  1.14 & & \\
\hline\hline
\end{tabular}
\end{table}
As it follows from the numerical results, the essential point in the
calculation of the Auger decay rates is to determine very fine features
of the wave function.
Really, the largest contribution to the transition matrix
element comes from the smallest component corresponding to the
largest possible $\lambda_1$ value (Table~\ref{tabaug3}) and
contributions to the
$M_{tr}$ from the ($l, \lambda_1$) components of $\Psi_{LN}$
compensate each other due to alternation of signs. Moreover, the
most important in the calculation of $M_{tr}$ is the large $\rho_1$
region of the configuration space, where these components decrease
exponentially.

As a result, uncertainty in the
calculated decay rates for the multipolarity $\lambda_0=3$ is of an order of
10 per cent. As is indicated in Table~\ref{tabaug}, the decay rates
for the $^{4}\! He\bar pe$ system in this case are in agreement
with the results of~\cite{mo}. The calculated decay rate of the
(34,4) state of this system is in fairly good agreement with
experimental lifetime $\tau$= 15ns~\cite{mor}. At the same time,
for the (33,3) state the calculated decay rate twice exceeds the
experimental value ($\tau$=16.6ns)~\cite{maas}.

On the contrary, the case of the transition multipolarity $\lambda_0=4$
is rather complicated for calculation due to much smaller values of
the transition matrix element, and only an estimate of the decay
rate can be obtained. These values exceed significantly the results
obtained in~\cite{mo}.

The calculated decay rates of the
$^{3}\! He\bar pe$ system reveal the substantial
isotopic dependence. In fact, decay rates of the (33,4) and (32,3)
states of $^{3}\! He\bar pe$ are about three times as large as
those of analogous (34,4) and (33,3)
states of $^{4}\! He\bar pe$. At the same time,
other characteristics of the ($L-1,N$) state
of $^{3}\! He\bar pe$ are close enough in comparison with the
($L,N$) state of $^{4}\! He\bar pe$. As for $\lambda_0=4$
transitions of $^{3}\! He\bar pe$, in this case the calculated
decay rates can be estimated only to an order of magnitude and
isotope effect is uncertain.

Thus, the Auger decay rates have been calculated for the
multipolarities $\lambda_0=3,4$. The method of calculation should
be improved to determine with more precision the small components
of the wave function and small decay rates in the cases of higher
multipolarities $\lambda_0 \geq 4$. Isotopic effect has been found
and the role of the structure of the wave function has been
studied.  The Auger decay rates of metastable states
$\lambda_0=3$ amount to tenth of the radiative transition rates and
can be important in cascade processes. If the isotopic effect
for the $\lambda_0=3$ states is the same as for $\lambda_0=4$
states, the Auger decay in $^{3}\!  He\bar pe$ can compete with
radiative transitions.

\section{Conclusions}
Properties of new three-body systems, the recently discovered
antiprotonic helium atoms and analogous hadron--containing systems
are intensively investigated both theoretically and
experimentally. The investigation of
intrinsic properties of antiprotonic helium atoms, their
formation and collisions with atoms and molecules is necessary for
understanding the antiproton fate in media.
The coexistence of a particle and an antiparticle in the
same atomic system brings on the principal possibility to study
antiproton properties by precise measurements of
antiprotonic helium atoms.

Nonstability and an extremely large
total angular momentum $L\sim 30-40$ of these exotic systems give
rise to significant difficulties in the theoretical treatment of
this system. However, due to very long lifetimes the metastable
states of these systems are treated as true bound states and the
variational approach is applied in the calculation.  Variational
eigenfunctions and eigenenergies are used to obtain the radiative
 transition rates, energy-level splitting due to relativistic
interactions and Auger decay rates of antiprotonic helium atoms.
The radiative transitions and Auger decay determine the cascade
process after the formation of antiprotonic helium. Calculation of
the energy-level splitting will be helpful in the proposed
measurement of the multiplet structure~\cite{pps}. One can mention
that these rather different values are sensitive to different
parts of the wave function, e.~g. the region of small interparticle
distances is important in the calculation of energy-level splitting
and the asymptotic region is important in the calculation of
Auger decay rates.

A simple set of trial functions used in the calculations,
nevertheless, provides a description of rather different
properties of antiprotonic helium atoms.
Convergence of the calculated values with increasing the
number of trial functions has been achieved and the isotopic
effects have been investigated by comparison of the calculated
properties of $^{4}\! H\! e\bar pe$ and $^{3}\! H\! e\bar pe$. In
this respect, one should mention the investigation of
$^{6}\!  H\! e\bar pe$ having in mind that the reduced-mass ratio
for $^{6}\! H\!  e\bar p$ and $^{4}\! H\! e\bar p$ is almost the
same as for $^{4}\! H\! e\bar p$ and $^{3}\! H\! e\bar p$.

The calculated wave functions can be used also to consider more
complicated problems; the most important of them is the
determination of formation probabilities and initial populations.
Due to substantial density and impurity effects found in
experiments~\cite{nak},~\cite{wid},~\cite{wid1}, other
important applications are the interaction of antiprotonic helium
atoms with usual helium atoms and diatomic molecules. As for the
Auger decay, more efforts are needed to calculate small
components of the wave function in the asymptotic region in
cases of higher multipolarities $\lambda_0 \geq 4$.

{\it Acknowledgement.}
The author is grateful to
J.~Eades, R.~Hayano, K.~Ohtsuki, I.~Shimamura, E.~Widmann and
T.~Yamazaki for useful discussions and to Computing and Network
Division of CERN for the valuable assistance in computations
necessary for these calculations.


\begin{thebibliography}{99}
\bibitem{yam} Yamazaki~T. e.a.,
Phys.  Rev. Lett., 1989, vol.~{\bf 63}, p.~1590
\bibitem{nakpi} Nakamura~S.~N. e.a.,
Phys. Rev.~A, 1992, vol.~{\bf 45}, p.~6202
\bibitem{iwa} Iwasaki~M. e.a.,
Phys. Rev. Lett., 1991, vol.~{\bf 67}, p.~1246
\bibitem{nak} Nakamura~S.~N. e.a., Phys.
Rev.~A, 1994, vol.~{\bf 49}, p.~4457
\bibitem{con} Condo~G.~T., Phys. Lett., 1964, vol.~{\bf ~9}, p.~65
\bibitem{rus} Russell~J.~E., Phys. Rev.~A, 1970,
vol.~{\bf 1}, p.~721; p.~735; p.~742
\bibitem{ahl} Ahlrichs~R., Dumbrajs~O., Pilkuhn~H. and
Schlaile~H.~G., Z.  Phys.~A, 1982, vol.~{\bf 306}, p.~297
\bibitem{mor} Morita~N. e.a., Phys. Rev. Lett., 1994,
vol.~{\bf 72}, p.~1180
\bibitem{hay} Hayano~R.~S. e.a., Phys. Rev. Lett., 1994,
vol.~{\bf 73}, p.~1485~(1994); vol.{\bf 73}, p.~3181
\bibitem{shim} Shimamura~I., Phys. Rev.~A, 1992, vol.~{\bf 46},
p.~3776
\bibitem{fb} Kartavtsev~O.~I.,
Few--Body Systems Suppl., 1995, vol.~{\bf 8}, p.~228
\bibitem{kar} Kartavtsev~O.~I., Proc. of the 3rd
Intern. Symp. on Muon and Pion Interactions with Media,
Dubna, 1995, p.~138
\bibitem{maas} Maas~F. e.a., Phys. Rev.~A, 1995, vol.~{\bf 52}, p.~4266
\bibitem{wid} Widmann~E. e. a., 1995, Proc. of the Third Conference
on Nucleon--Antinucleon Physics, Moscow, Yad. Fiz., (in print)
\bibitem{kor} Korobov~V.~I., Proc. of the Intern. Symp. on
Muon Catalyzed Fusion, Dubna, 1995, Hyperfine Interactions (in
print)
\bibitem{mo} Morita~N., Ohtsuki~K. and Yamazaki~T.,
Nucl. Instr. Meth.~A, 1993, vol.~{\bf 330}, p.~439
\bibitem{yamoht} Yamazaki~T. and Ohtsuki~K.,
Phys. Rev.~A, 1992, vol.~{\bf 45}, p.~7782
\bibitem{eades} Charlton~M., Eades~J., Horvath~D.,
Hughes~R.~O., Zimmermann~C., Phys. Rep. 1994, vol.~{\bf 241}, p.~65
\bibitem{pps} PS205 Collaboration, 1995, Preprint CERN
SPSLC 95--12/SPSLC I~201
\bibitem{exan} Kartavtsev~O.~I., Proc. of the
International Symposium on Exotic Atoms and Nuclei, Hakone, Japan,
1995, Hyperfine Interactions (in print)
\bibitem{lev} Lev~F.~M., Riv. Nuova Cim. 1993, vol.~{\bf 16}, p.~1
\bibitem{aug} Fedotov~S.~I., Kartavtsev~O.~I., Monakhov~D.~E.,
1995, Proc. of the Third Conference on Nucleon--Antinucleon
Physics, Moscow, Yad. Fiz. (in print); JINR Rapid Communications,
1995, No.~5~(73), p.~13
\bibitem{wid1} Widmann~E. e. a., Phys. Rev.~A, 1995, vol.~{\bf 51},
p.~2870
\end{thebibliography}
\end{document}